\documentclass[12pt, letterpaper]{article}

\usepackage{amsmath,amssymb}
\usepackage{cite}
\usepackage{fancyhdr}
\usepackage[top=1in, bottom=1in, left=1in, right=1in]{geometry}
\usepackage[dvipdfm, dvips]{graphicx}
\usepackage{hyperref}

\numberwithin{equation}{section}

\begin{document}


\setcounter{page}{0}
\date{}

\lhead{}\chead{}\rhead{\footnotesize{RUNHETC-2011-??\\SCIPP-11-01}}\lfoot{}\cfoot{}\rfoot{}

\title{\textbf{Fuzzy Geometry via the Spinor Bundle, with Applications to Holographic Space-time and Matrix Theory\vspace{0.4cm}}}

\author{Tom Banks$^{1,2}$ and John Kehayias$^{2}$\vspace{0.7cm}\\
{\normalsize{$^1$NHETC and Department of Physics and Astronomy, Rutgers University,}}\\
{\normalsize{Piscataway, NJ 08854-8019, USA}}\vspace{0.2cm}\\
{\normalsize{$^2$SCIPP and Department of Physics, University of California,}}\\
{\normalsize{Santa Cruz, CA 95064-1077, USA}}}

\maketitle
\thispagestyle{fancy}

\begin{abstract}
\normalsize \noindent
We present a new framework for defining fuzzy approximations to geometry in terms of a cutoff on the spectrum of the Dirac operator, and a generalization of it that we call the Dirac-Flux operator. This framework does not require a symplectic form on the manifold, and is completely rotation invariant on an arbitrary n-sphere.  The framework is motivated by the formalism of Holographic Space-Time (HST), whose fundamental variables are sections of the spinor bundle over a compact Euclidean manifold.  The strong holographic principle (SHP) requires the space of these sections to be finite dimensional.  We discuss applications of fuzzy spinor geometry to HST and to Matrix Theory.
\end{abstract}


\newpage
\tableofcontents
\vspace{1cm}


\section{Introduction: Holographic Space-time (HST)}

HST is an attempt to supply a general formalism for a theory of quantum gravity, which will reduce to string theory for space-times that are asymptotically AdS or Minkowski, but which has the flexibility to discuss cosmology, including dS space.  The formalism also makes more direct contact with concepts of local physics than any extant string theoretic formalism.
The Strong Holographic Principle (SHP), originally stated by Bousso and much championed by TB and W.~Fischler is the assumption that the Covariant Entropy Bound (CEB)\cite{ceb_fs, ceb_b1, ceb_b2} implies that the Hilbert space encoding all measurements inside a causal diamond, is finite dimensional, with dimension that approaches the exponential of one quarter of the area of the holographic screen of the diamond\footnote{The boundary of a causal diamond is a null surface, which can be foliated by space-like surfaces. The holographic screen of the diamond is the space-like surface of largest area.  We abuse language and call the area of the screen the area of the diamond.}.  The area is measured in Planck units and the formula is supposed to be only asymptotic for large area.   In weakly coupled string theory, there is a further caveat.  Here the Einstein equations break down at a length scale parametrically larger than the Planck scale and the identification of entropy and area fails unless the area is large in string units. The SHP combined with the notion of commutativity at space-like separation, encodes all of the geometrical properties of a Lorentzian space-time into quantum mechanical statements about operator algebras.

The basic idea is that space-time is only an emergent phenomenon, but that its properties reflect more basic properties of the underlying quantum theory of gravity.   The kinematics of HST is a net of finite dimensional operator algebras, called diamond algebras $A(D)$, with specified intersections $O(D,D^{\prime})$, which are tensor factors in both $A(D)$ and $A(D^{\prime})$. $O(D,D^{\prime})$ represents the set of all quantum measurements, which can be performed in the maximal area causal diamond in the intersection of the diamonds $D$ and $D^{\prime}$. 
In quantum field theory (QFT) in a fixed space-time, the diamond algebra $A(D)$ is constructed from fields smeared with test functions,  whose support lies within the diamond $D$. Algebraic quantum field theory is a formulation of QFT in terms of the {\it net} of diamond algebras and their intersections\cite{aft}.  The causal structure of space-time is completely encoded in the structure of the net of operator algebras.

In QFT, as a consequence of conformal invariance at short distances, the operator algebras are all infinite dimensional.  HST postulates instead that the algebras for diamonds with finite area holographic screens are finite dimensional matrix algebras, operating in a Hilbert space whose dimension is the exponential of one quarter of the area (in Planck units) of the screen.  We can turn this around and say that it is a {\it definition} of the area of the screen in terms of purely algebraic properties of the net of algebras. Thus, in HST, both the conformal factor and the causal structure may be defined in terms of properties of the quantum operator algebra.

It is clear that specifying the data in these algebras for a sufficiently rich set of diamonds, in the limit in which space-time emerges, will determine both the conformal factor and the causal structure of the Lorentzian geometry, which are thus kinematical properties of the quantum theory, rather than fluctuating quantum variables. The arguments of Jacobson\cite{ted}, suggest that Einstein's equations for the geometry will be an automatic consequence of the laws of thermodynamics, in the emergent space-time limit. Indeed, Jacobson argued that {\it any} theory, which obeyed the laws of thermodynamics, and for areas large in Planck units, obeyed the Bekenstein-Hawking area/entropy law for areas transverse to each local Rindler horizon, would satisfy Einstein's equations.

The actual quantum variables may be thought of classically as the space-time orientations of pixels on the holographic screen\cite{susyholoscreen}.  Naively, a pixel is a position on the screen, through which a null ray passes, and the orientation of a bit of $d-2$ plane orthogonal to the null ray.  This data is incorporated in the Cartan-Penrose (C-P) equation

$$\overline{\Psi} \gamma^{\mu} \Psi (\gamma_{\mu} )_{\alpha\beta} \Psi_{\beta} = 0,$$ which forces the vector bilinear to be null, and the spinor $\Psi$ to be a null plane spinor for that null ray: $$\Psi = \begin{pmatrix} 0 & S_a \end{pmatrix}.$$  The C-P equation is Lorentz covariant and has a scaling symmetry.  These are considered gauge equivalences.  Generically, we may expect them all to be fixed in a unitary formulation of the quantum mechanics (which is all that we will consider).  In fact, the scaling symmetry is explicitly broken in the quantum theory. However, there is a $\mathbb{Z}_2$ subgroup of the scaling symmetry, which is preserved and ends up playing the role of the $(-1)^F$ gauge symmetry familiar from quantum field theory. The connection between spin and statistics is automatic, as in Matrix Theory\cite{bfss}. The Lorentz gauge symmetry is fixed by insisting that the direction of the null vector is determined by the coordinate on the holographic screen.  For example, for a spherical screen, parametrized by a $d-1$ dimensional unit vector ${\bf \Omega}$, the null vector is $(1, {\bf \Omega})$.  The solution of the C-P equation on each infinitesimal pixel on the screen is a section of the spinor bundle over the holographic screen.

The strong holographic principle implies that a finite area holographic screen can have only a finite number of pixels, and that the algebra of variables for each pixel has a finite dimensional unitary representation.  A finite number of pixels means a finite basis of sections of the spinor bundle over the holographic screen.  The purpose of this paper is to propose a definition for this {\it fuzzy spinor bundle}.  The screen is a compact Riemannian manifold and the Dirac operator of the screen has a discrete unbounded spectrum.  A sharp cutoff on the Dirac eigenvalue gives a finite dimensional approximation to the spinor bundle, and, as we shall see, this provides a new definition of  fuzzy geometry.

 For compactifications to four dimensional space-time, the quantum commutation relations take the form

$$[(\psi^P)_i^A , (\psi^{\dagger\ Q})^j_B ]_+ = \delta_i^j \delta^A_B Z^{PQ} .$$  The indices $i,j$ run from $1$ to $N$, $A,B$ run from $1$ to $N+1$, and $P,Q$ run over a basis of a finite dimensional approximation to the spinor bundle over a compact internal manifold, obtained by cutting off the Dirac spectrum on that manifold. We call this the {\it pixel algebra} of the HST model. It must be supplemented by commutation relations between the $Z^{PQ}$ and the fermionic variables, forming a finite dimensional super-algebra.  The holographic principle implies this algebra must have a finite dimensional unitary representation.  We assume further that the action of the fermionic operators sweeps out the entire space of states of this representation.   In writing this equation we have anticipated a property of the Dirac equation on the two sphere, namely that an eigenvalue cutoff is equivalent to an angular momentum cutoff.   The space of $N\times N+1$ matrices contains all spinor spherical harmonics up to spin $N- \frac{1}{2}$, and as $N \rightarrow\infty $ it approximates the chiral spinor bundle over the two sphere, while the space of Hermitian conjugate matrices converges to the anti-chiral spinor bundle.
This elegant choice for the finite dimensional approximation, is based on A.~Connes ideas\cite{ncg} about non-commutative geometry \cite{steinacker}. As noted,
$\psi$ and $\psi^{\dagger}$ are the two chiral spinor bundles over the fuzzy two sphere. $Z^{PQ}$ lives in the bundle of forms over the compact internal manifold, fuzzified as the product of two cutoff spinor bundles. The $Z^{PQ}$ are the analogs of wrapped brane charges in string theory.  We will call the finite dimensional irreducible representation of this super-algebra ${\cal P}$, and refer to it as {\it the pixel Hilbert space}.

For the two sphere, our fuzzification is the same as that defined by Berezin quantization\cite{berezin}, which is a special case of the geometric quantization of symplectic manifolds.  Berezin noted that the spaces of sections of holomorphic vector bundles on a Kahler manifold are finite dimensional.  Holomorphic vector bundles are carry a $U(1)$ gauge connection, and the associated fluxes through two cycles are integers characterizing the bundle.  As we take the fluxes to infinity, the dimension of the space of sections goes to infinity as well, and if the bundle is {\it ample}, the set of functions
$$F(M) = \sum \bar{s}_i (\bar{z} ) M^i_j s^j (z) ,$$ becomes a basis in the algebra of functions on the manifolds.  Here $M$ is a general complex matrix acting on the space of sections, and the question of how
smooth the functions are depends on the behavior of the matrix elements in the limit.  For the two sphere, there is an $SO(3)$ symmetric Kahler form, the volume form in the round metric, and the quantum number which characterizes vector bundles is essentially the angular momentum.  The Dirac equation is spherically symmetric and an eigenvalue cutoff is the same as an angular momentum cutoff, so the two procedures are equivalent.  On higher dimensional spheres there is no spherically symmetric Kahler form nor even a Poisson structure that is spherically symmetric.  By contrast, the Dirac eigenvalue cutoff is spherically symmetric on all spheres, so it is inequivalent to any fuzzification based on the ideas of geometric quantization.   Many alternative fuzzifications of higher dimensional spheres have been proposed (see for instance \cite{fuzzynspheres1, fuzzynspheres2, fuzzynspheres3}), but they are all much more complicated than the simple Dirac cutoff.

There are two important points to made about our choice of Dirac fuzzification:
\begin{itemize}
\item Our intent is to provide a general tool for HST, which is a particular approach to quantum gravity.  The fundamental variables in HST are a finite dimensional approximation to the space of sections of the spinor bundle over the holographic screen.  No other definition of fuzzy geometry gives a simple description of spinor bundles for general geometries.
\item Our approach does not exactly fall into the conventional {\it spectral triple} classification\cite{ncg}, since, as we will see below, the natural algebra which arises in the geometric limit is the algebra of bounded operators on the space of square integrable sections of the spinor bundle.  The usual algebra of functions, the focus of most approaches to fuzzy geometry, is a proper subalgebra of this, as is the algebra of forms (with Clifford, rather than Grassmann multiplication), but the natural limit of the non-commutative fuzzy algebra is the full non-commutative operator algebra.
\end{itemize}

The approach to geometry via the cutoff Dirac operator contains all of the geometrical information about a compact Riemannian manifold (in the appropriate limit) so it seems to us that it is an interesting definition even if it does not fit exactly into the framework of previously proposed axioms.

\subsection{Cosmological, Asymptotically Flat, and Asymptotically AdS Space-times}

This subsection has little to do with the rest of the paper and can safely be skipped.  We include it at the behest of the referee, who felt that the general framework of HST was sufficiently unknown that a bit of explanation would be helpful to most readers.  The operator algebras of HST and QFT differ only in the finiteness of causal diamond algebras for finite area diamonds.  However, the Hamiltonian formulation of the two frameworks is completely different.  In HST, the entire Hilbert space of the theory is associated with a {\it single } time-like trajectory, and operators in that Hilbert space refer to actual observations made by an observer following that trajectory.  For example, the description of the planet Saturn, for an observer that spends all of history on Earth, is in terms of photons that have made the journey to Saturn and been reflected back to earth.  HST becomes a theory of space-time because it has an infinite number of other descriptions in terms of other time-like trajectories, and a consistency requirement that physics accessible to two different observers has two unitarily equivalent descriptions.  We will label the different time-like trajectories by a discrete parameter ${\bf x}$, which lies in a lattice whose topology is that of flat space in some number of dimensions. 

A time-like trajectory at ${\bf x}$ is encoded into QM, using the strong holographic principle, by specifying a sequence of Hilbert spaces ${\cal H} (n, {\bf x}) = \otimes {\cal P}^{n(n+1)} $, where we are restricting attention to $4$ non-compact dimensions\footnote{We include dS space in our definition of non-compact geometries, despite the fact that in global coordinates it has compact spatial sections.  The full dS space is really the thermo-field double of a single horizon volume\cite{susskindetal}, and the latter is a non-compact manifold because it has a boundary.}.  The geometry of the compact dimensions is encoded in ${\cal P}$, and this fact is the primary burden of the present paper.  These spaces correspond, via the holographic principle, to larger and larger segments of the time-like trajectory.  For small causal diamonds, $n$ is proportional to the proper time, in Planck units.  Among maximally symmetric spaces, different values of the cosmological constant imply different behaviors of the maximal value of $n$ as the proper time goes to infinity.  In AdS space, $n$ becomes infinite in a finite proper time of order the AdS radius.   In asymptotically flat space, the two go to infinity at fixed ratio, while in dS space the maximal value of $n$ is finite and proportional to the dS radius in Planck units.  The evolution operator $U(t,0)$ or $U(t,-t)$ (for a Big Bang, or TCP symmetric space-time, respectively, $t$ is proportional to $n$) operates in the full Hilbert space ${\cal H} (n_{max}, {\bf x})$, but for any $t$, it factorizes into a product $U_{in} \otimes U_{out}$, where $U_{in}$ operates only in ${\cal H}_n$ for the causal diamond whose tips are labelled by $(t,0)$ (in the Big Bang case), and $U_{out}$ operates in the tensor complement of ${\cal H}_n$ in ${\cal H}_{n_{max}}$.  If $n_{max}$ is infinite, the tensor complement must be defined by a careful limiting procedure.   This rule incorporates causality into the formalism.  The dynamics of the variables inside some causal diamond does not depend on those outside it.  The reader should be aware that the definitions are all purely quantum mechanics. The geometrical picture is emergent, via the holographic principle, in the large diamond limit.

The quantum system defined by these rules is {\it a complete description of the universe as viewed by a given observer}.  The rest of space-time is, in HST, a gauge redundancy, but gauge invariance puts very strong constraints on the single observer dynamics. To define an HST, we assign evolution operators $U(t,0; {\bf x})$ to each trajectory in our infinite congruence.  We then specify overlap Hilbert spaces
${\cal O} (n, {\bf x,y})$, which, for each $n$, are tensor factors in both ${\cal H}_n (\bf x)$ and $  {\cal H}_n (\bf y) $.  For nearest neighbors on the lattice ${\cal O} (n, {\bf x,y}) = {\cal H} (n, {\bf x}) = \otimes {\cal P}^{n(n-1)}$, and more generally, its dimension is a decreasing function of $d({\bf x,y})$, the minimal number of lattice steps between the two points.   Starting from some initial state, for each observer, time evolution will produce two sequences of density matrices $\rho (n, {\bf x})$ and $\rho (n, {\bf y})$.  For every $n$ and ever pair of points, these two density matrices must be unitarily equivalent to each other.
In words we say that ${\cal O}(n, {\bf x,y})$ contains the information that is accessible to both observers at time $n$.  Geometrically, we can view it as the Hilbert space in the maximal causal diamond, which fits into the intersection between the diamonds of the two trajectories at the time $n$. The consistency condition says that up to a unitary change of basis, the two observers predict the same density matrix for all of this common information, at all times.  It constrains both the mutual time evolutions and the mutual choices of initial state.   This infinite set of conditions is rather hard to satisfy, and so far only a few consistent models have been found.  They seem to describe both the very beginning and the very end of cosmological history (assumed to be a dS space), at which times all of the degrees of freedom are in thermal equilibrium.  These are regimes in which conventional QFT descriptions of physics are most at odds with the true behavior of the models, because QFT gets the thermodynamics of gravitational systems wrong.

There is also a partial understanding of how to use the HST formalism to recover conventional string/M-theory descriptions of space-times which are asymptotically flat or AdS.  The basic idea is that in those space-times, one should restrict attention to the largest causal diamonds, which is to say the conformal boundary.  This means that $N \rightarrow \infty$ in the quantum super-algebra, while  the number of basis sections of the spinor bundle over the internal manifold is kept fixed if its dimensions are fixed in Planck units.   The resulting smooth 2-sphere has infinite radius and the theory must become conformally invariant in order for observables to have finite limits.  In asymptotically AdS space, the conformal boundary is $R \times S^2$, and the relevant conformal group is $SO(2,3)$.  Well known arguments, which go under the rubric AdS/CFT correspondence, imply that the limiting theory must be a conformally invariant QFT on $R \times S^2$.   We have a lot of experience constructing QFTs as limits of systems with a large number of variables, each of which has a finite dimensional representation space, so the mapping of the HST formalism into AdS/CFT in the large $N$ limit seems plausible, though none of the details has been worked out.   

For asymptotically flat space, the conformal boundary is null and the only relevant conformal group is $SO(1,3)$ , which is interpreted as the Lorentz group.  If the internal manifold has a covariantly constant spinor, then it has a Dirac zero-mode, which is preserved by our definition of fuzzy geometry.  The corresponding scalar fermion bilinear is just a constant function on the internal manifold, so a subset of the anti-commutation rules read
$$[(\psi^0)_i^A , (\psi^{\dagger\ 0})_B^j ]_+ =  \delta_i^j \delta_B^A .$$  In the limit when $N\rightarrow\infty$, a singular basis of spinor spherical harmonics are delta function measures on the sphere, multiplied by constant spinors.  The conformal Killing spinor equation on the sphere has a solution space which transforms as the $(2,1) \oplus (1,2)$ representation of $SO(1,3)$. When these are integrated against the delta functions we obtain operators satisfying\cite{bfm}
$$[ Q_{\alpha} ({\bf \Omega}), Q^{\dagger}_{\dot{\beta}} ({\bf \Omega})]_+ = K (1, {\bf \Omega})_{\mu} ( \sigma^{\mu})_{\alpha\dot{\beta}}, $$ where we have used the usual Weyl four vector of two by two matrices and $K$ is a positive normalization constant, which arises when taking the limit.  Indeed \cite{bfm,bfss}, we can take a more general limit, with block diagonal matrices of sizes $N_i$, all of which go to infinity at fixed ratio, and obtain a Fock space of massless superparticles with all possible momenta.  Thus, kinematically, we can obtain a limit of the HST system corresponding to a massless supersymmetric field theory, whenever the internal manifold has a covariantly constant spinor. Notice that the momenta in the Poincare algebra arises as an auxiliary bilinear in the underlying fermionic variables, and we only obtain a Poincare invariant limit as a consequence of SUSY.

This has been only a sketch of the HST formalism.  More details can be found in \cite{HSTref1, HSTref2, HSTref3}.  The purpose of the present paper is to present the definition of fuzzy geometry which underlies this theory.

\section{The Dirac equation and geometry}

Alain Connes\cite{ncg} has made the Dirac operator the central focus of his metrical non-commutative geometry \cite{steinacker}.  Connes emphasis is on non-commutative geometries with infinite dimensional function algebras, while we are concerned with finite dimensional non-commutative approximations to ordinary commutative geometries.  For physicists, an easy way to understand the relation between the Dirac equation and geometry is to think about the short time expansion of the heat kernel for the square of the Dirac operator

$$\langle x | e^{- t D^2} | y \rangle \rightarrow K\ t^{-\frac{d}{2}} e^{- \frac{l^2 (x,y)}{4t}} ,$$ where $d$ is the dimension of the manifold and $l(x,y)$ the geodesic distance between the points.  The factor $K$ is the number of geodesics of equal minimal length connecting the two points.  This expression is most easily derived from the Feynman path integral representation of the heat kernel. The short time limit is a semi-classical limit for that functional integral. The heat kernel thus contains all of the geometrical information about the manifold.

Note that for this expression we need to know not only the spectrum of the Dirac operator, but also the form of its eigensections in the position representation.  Geometers have long known how to describe the points of a manifold in terms of the algebraic structure of its algebra of functions. A point is equivalent to the maximal ideal of functions which vanish at that point. Alternatively, a point defines an algebra homomorphism between the algebra of functions and the complex numbers (a multiplicative linear functional).  Connes shows that everything that is to be known about a manifold can be encoded in the relation between the Dirac operator and the algebra of smooth functions realized as multiplication operators on the Hilbert space of square integrable sections of the spinor bundle.  He then proposes an abstract definition of the Dirac operator for a general non-commutative algebra of operators on a Hilbert space as the definition of a non-commutative Riemannian manifold.

Our aim is more modest. We simply want to recover the normal commutative geometry of manifolds as a limit of finite dimensional matrix algebras. This is relatively straightforward.  For most\footnote{To quantify the notion of most, we have to think about a moduli space of Riemannian manifolds satisfying some equations.  Such moduli spaces have a natural metric on them, and although non-compact, the moduli space has finite volume.  This means that extreme values of the moduli are ``non-generic".  Our statement will be valid in a region of moduli space that contains a large fraction of the total volume.} compact Riemannian manifolds of dimension $d$ and volume $V$, the operator $V^{\frac{1}{d}} D$ has a spectrum that runs from $\sim \pm 1$ to $\pm\infty$.   We will define a fuzzy spinor bundle over this manifold by cutting off the spectrum of this operator via the inequality
$|| V^{\frac{1}{d}} D || < N $, where $N$ is a positive integer. That is, we restrict to the space of eigensections whose eigenvalues satisfy this inequality.  The dimension of this subspace of eigensections is another positive integer $K(N)$.

The algebra of $K(N) \times K(N)$ matrices is realized as a set of integral kernels $$M_{\alpha\beta} (x,y) = \sum  M_{ij} \psi_{\alpha}^{*\ i} (x) \psi_{\beta}^j (y),$$ on the full spinor bundle.  In the limit $N\rightarrow\infty$, we can restrict attention to matrices which produce kernels of the form
$$\sum  M_{ij} \psi_{\alpha}^{*\ i} (x) \psi_{\beta}^j (y) \rightarrow f_{\alpha\beta} (x) \delta (x,y),$$ where $\delta (x,y)$ is the Dirac distribution on the manifold.  $f_{\alpha\beta}$ belongs to the algebra of differential forms with Clifford multiplication, rather than the standard Grassmann product.  The Clifford multiplication of course depends on the metric.  With appropriate restrictions on the limiting form of $M_{ij}$ we can get measurable, continuous, or smooth differential forms.  However, there is no particular reason to do this in the spinor formalism.   The general theory of approximating bounded operators on a Hilbert space by operators of finite rank, leads us to consider the full non-commutative algebra of bounded operators on the space of square integrable sections of the spinor bundle as the natural algebra of the continuous geometry.
This contains the algebras of functions and differential forms as subalgebras, and is no less of a characterization of the geometry of the manifold than those more familiar ones.

\subsection{Moduli}
If we have a moduli space of manifolds, then the eigenvalues and eigensections of the Dirac operator depend smoothly on the
moduli.  However, the spirit of non-commutative geometry \cite{steinacker} and fuzzy geometry in particular is that the algebra determines the geometry.  In the standard geometric quantization of the two torus, we can see that this leads to a discretization of moduli space.
A square fuzzy torus is defined by the algebra of all $N \times N $ matrices, written in terms of generators $U,V$ satisfying
$$U^N = V^N = 1,$$ $$ UV = e^{\frac{2\pi i}{N}} VU.$$   The area of this torus in Planck units is $\sim N^2$.  If $N$ has a factor $k$, we can get a rectangular torus by restricting attention to the subalgebra generated by $U^k$ and $V$, and a similar restriction produces tilted tori as well.  But we only get a rational set of moduli in this manner.   Continuous moduli arise, like longitudinal momenta in Matrix Theory\cite{bfss} and HST, as ratios of integers, both of which are taken to infinity.

For spinor fuzzification we consider the Dirac operator with periodic boundary conditions\footnote{The implications of different spin structures for our program seem interesting, but we have not understood them.}.  A general 2-torus is determined by a parallelogram, parameterized in terms of three real numbers $(a,b,c) $ with $0 < c < a$.  $a$ is the length of the horizontal segments, and $b$ the vertical separation between them.   $c$ determines the tilt of the parallelogram.   The eigenvalues and eigensections of the Dirac operator with periodic boundary conditions are determined by a two vector ${\bf p} = (p_1 , p_2)$
with $$ p_1 = \frac{2\pi n}{a}\  \  \  \  p_2 = \frac{2\pi m}{b} - \frac{2\pi nmc}{ab}.$$  The eigenvalues are $\pm |{\bf p}|$ and the eigensections are $$\psi_{\pm} e^{i {\bf p\cdot x}},$$ where $\psi_{\pm}$ are the two eigenspinors of $\sigma_1 p_1 + \sigma_2 p_2$.

Fuzzification consists of choosing integer valued moduli $a = N$, $b = M$, $c = k \leq N$ and cutting off the values of $m $ and $n$.  Two natural cutoffs are $n \leq N$, $m \leq N$, and $\left(\frac{n}{N}\right)^2 + \left(\frac{m}{M} - \frac{knm}{MN}\right)^2  < K^2, $ for some integer $K$.  The first is similar to the kind of cutoff one gets from Kahler quantization, while the latter conforms to our general idea of just bounding the spectrum of the Dirac operator.   For $K$ of order $1$, both methods give a number of sections of the spinor bundle that scales like $MN$, which is proportional to the area of the torus.  If we make the independent sections into independent generators of a quantum superalgebra, then the entropy of the torus will be proportional to its area.

More generally, the large eigenvalues of the Dirac equation on any smooth compact manifold are approximately like plane waves and their degeneracy grows like $P^D$, where $P$ is the eigenvalue cutoff and $D$ the dimension.  Thus, the number of independent sections grows like the volume of the manifold in Planck units.  Since this compact manifold is the holographic screen of a Lorentzian manifold in the HST formalism, this is precisely the right Bekenstein-Hawking entropy in the general case.  That is to say, the entropy per four dimensional pixel (fixed value of $i$ and $A$) will, for compact dimensions large in higher dimensional Planck units, be proportional to the volume of the internal dimensions.  This is the conventional Kaluza-Klein relation between the four dimensional and higher dimensional Planck scales.

It is easy to work out the spinor fuzzification of a general torus, and we will do a general sphere in the next section.   The procedure is straightforward for any manifold for which one can work out the eigenvalues and eigensections of the Dirac equation. For a general torus, the eigenvalues are determined by a D dimensional lattice of momenta, with metric $G_{ij}$, and the eigenvalue cutoff is $$P^i G_{ij} P^j < N^2,$$ which is satisfied by only a finite set of momenta on the lattice.  The eigensections are exponentials multiplied by constant spinors.  For large $N$, the eigenvalue degeneracy scales like $N^D$.  Again it is clear that as we make continuous changes in the moduli of the torus, the number of eigenvalues satisfying the bound jumps discretely.  Thus, we can consider a discrete set of moduli, {\it e.g.}~at the boundary of these jumps.  The HST formalism writes anti-commutation relations for a finite set of eigensections, and the fact that the resulting superalgebra has a finite dimensional unitary representation means that the physics for a fixed number of eigensections has no continuous parameters.  Of course, as $N$ gets large the discrete set of moduli become dense in moduli space, and we recover the familiar properties of continuous geometry.

From the point of view of approximating geometry, we may view the restriction to rational moduli as a convenience only.  That is, we could look at finite dimensional approximations to the spinor bundle for general, continuous values of the moduli.
The eigenvalues and eigensections of the Dirac equation depend continuously on the moduli.  However, in the context of HST, the spinor eigensections become quantum operators, and generate a superalgebra with a finite dimensional unitary representation.  All of the physics of the HST models depends only on this representation, and there are no continuous parameters.  One comes to the conclusion that the continuous moduli of conventional string theory are the result of approximations in which some length scale is infinitely larger than the Planck scale at zeroth order.

The tensor product relation between spinor bundles and the bundles of differential forms imply that some of the topological features of the manifold are encoded in zero modes of the Dirac equation.  This is familiar from the Atiyah-Singer Index theorem and its generalizations.  In particular, if we have a covariantly constant spinor, $D_{\mu} \psi_0 = 0$, then it is also a zero mode of the Dirac equation.  The non-vanishing differential forms
$$ \overline{\psi}_0 \gamma_{\mu_1 \ldots \mu_k} \psi_0 , $$ where the matrices are the k-fold anti-symmetrized products of tangent space Dirac matrices, contracted into the vielbein, are all elements of the cohomology of the manifold.  This part of the topological information about the manifold is preserved by spinor fuzzification. Note that this is a bit different than Kahler fuzzification, where the information that is kept is a cutoff version of the Picard group and the dimensions of spaces of sections of holomorphic line bundles, as well as information about the complex structure.   It is peculiar though that not all of this information is invariant information about the finite dimensional matrix algebra.   For example the fuzzy square torus and the fuzzy sphere have the same algebra, and in some sense are distinguished only by the choice of a basis in this algebra (spherical harmonics vs. powers of clock and shift operators) and the way in which expansion coefficients in these bases behave in the large $N$ limit.

We believe that the lack of some explicit topological information about the manifold in fuzzy quantization is at the root of string dualities.   Highly supersymmetric compactifications of string/M theory to asymptotically flat space-times are often characterized by moduli spaces of classical background geometries.  The use of classical backgrounds that are solutions of some low energy effective field theory always implies that we are working in a limit where some length scale is much larger than the Planck scale\footnote{This can be a geometric length scale in the compactification manifold or the Compton wavelength of some quantum excitation.}.  We've seen that in such limits, the discrete moduli spaces of fuzzy compactification give rise to continuous ratios of large integers\footnote{For example, in Kahler quantization, the Kahler moduli have to do with the direction in the Picard group in which we take fluxes to infinity at fixed ratio.  Complex structure moduli have to do with choices of subalgebras of the algebra of all $N\times N$ matrices in the space of sections of the holomorphic line bundle corresponding to the chosen Picard group element. We've seen in the example of the two torus, that such sub-algebras are characterized by rational fractions $\frac{k}{N}$, where $k$ is a divisor of $N$.  These parameters become continuous as $N\rightarrow\infty$.  The example of tori shows how a similar phenomenon arises for spinor fuzzification. The number of Dirac eigenvalues below some bound is an integer, and jumps at discreet points in torus moduli space.  We can cover all possibilities in the $N\rightarrow\infty$ limit, by choosing rational values for the moduli with a maximum denominator of order the bound $N$.}.  The notion of continuous moduli spaces is conceptually wrong, but valid to all orders in expansions in $\frac{L_{P}}{L_{Large}}$.   String duality relations are derived in terms of constraints on low energy Lagrangians in two different limits, which have the same SUSY algebra.

In HST, the SUSY algebra arises in the limit of large causal diamonds in the non-compact space, with the discrete internal moduli fixed.  In that limit, the pixel algebra generators become distributions, $(\psi^P)_i^A \rightarrow \psi^K \delta (\Omega, \Omega_0) $ and the anti-commutation relations become (for $4$ dimensional asymptotically flat space)
$$[\psi^K , \psi^{\dagger L} ]_+ =  P Z^{KL}.$$ Recall that $K$ and $L$ label a finite dimensional basis of the space of Dirac eigensections on the internal manifold, with eigenvalue less than some bound.  $P$ is a positive real number.  It arises as follows. We take the $N$ characterizing the maximal spherical harmonic in the pixel algebra to infinity, obtaining wave functions localizable on the sphere, which deserve to be called particles penetrating the holographic screen.  Now we can do this in block diagonal matrices of size $N_i \rightarrow \infty$, with $\frac{N_i}{N_j} $ fixed, obtaining continuous longitudinal fractions.  We now view these fractions as ratios of dimensionfull momenta, and $P$ is that momentum.  If the internal manifold has a covariantly constant spinor, then we smear the distributional pixel algebra generators with conformal Killing spinors on the two sphere and pick $K,L$ to both be the zero mode, we get the ${\cal N} =1$ SUSY algebra with 4-momentum $P_{\mu} = P(1, \pm \Omega_0 )\footnote{ The $\pm $ ambiguity arises from a reflection ambiguity in the conformal Killing spinor equation.  It has to do with incoming and outgoing particles, and we will not discuss it further.} $.  

The pixel SUSY algebra will have scalar charges corresponding to BPS states if the theory has larger supersymmetry or more non-compact dimensions.  However, if the internal manifold has finite volume in Planck units\footnote{Translation: the representation space of the pixel algebra has finite dimension for fixed $N$.} then the eigenvalues of the charge operators are bounded.   It's easy to see that the bound corresponds to the point at which a state of that charge has a mass larger than the 4D Planck mass, so that it is really a black hole.  Such black holes can be made in particle collisions.   It is only in extreme limits of the discrete moduli, where the dimension of the pixel algebra goes to infinity, that we can describe ``all" of these black hole states as elementary objects like D-branes or Kaluza-Klein modes of compactified particles.   Indeed, such limits are always characterized by a small dimensionless parameter $g^2$. The non-gravitational nature of the states is only valid for values of charge less than some inverse power of $g^2$.  

The upshot of this discussion is that we know how to describe SUSY algebras and BPS states in the HST formalism.  A dual string pair corresponds to taking two different limits of the discrete parameters that characterize an HST compactification, namely the pixel algebra.  We can follow states between the two limits by following their conserved charges. In the two limits, the moduli become continuous parameters and we can use the usual arguments to compare the dual formulations of the theory.
One of us (TB) has been guilty on many occasions of saying that dualities proved that there were lengths smaller than the Planck scale in string theory (since {\it e.g.}~the weak coupling IIA string limit is a zero radius circle in M-theory).  This argument is specious.  Every calculable limit of string moduli space, as well as limits like F-theory, which are only partially calculable, depends on having a length scale much larger than the Planck scale of the non-compact dimensions, defined by the Einstein frame Lagrangian.  The expansion parameter is always a power of this ratio of scales.  This is the reason that the constraints of the Holographic Principle and the fundamentally discrete nature of moduli are not apparent in these expansions.

The discreteness of moduli has profound implications for cosmology. Much of the literature on string inspired cosmology, including many papers written by one of the authors (TB), uses moduli fields as ingredients in an inflationary cosmology.  Coherent fields are, from the HST point of view, an approximate way of describing states with many particles.  However, the particle horizon at early times is small, and the HST formalism only admits a finite number of particles in such a region.  The entropy of the particle horizon in a pre-inflationary era is roughly $$S = \frac{K}{\rho} \sim N^2 V_I ,$$ where $K$ is a geometrical factor that depends on the details of the early history of the universe, and $\rho$ is the energy density in Planck units.   $V_I$ is the number of independent sections in the fuzzy spinor bundle over the internal space, and $N^2$ is the number of spinor harmonics on the fuzzy two sphere.   When we make multi-particle states using the HST variables, the number of particles scales like $N^{1/2}$ if we require the particles to be roughly localizable\footnote{In order to use the conventional field theory calculation of inflationary fluctuations, we have to consider particles that are localizable on a scale much smaller than the horizon.}. For unification scale inflation we have $S \sim 10^{12} $ at most.  Thus, the number of particles is of order $$10^3 V_I^{-\frac{1}{4}} .$$  Thus, the $V_I \rightarrow \infty$ limit in which the internal geometry has approximately continuous moduli, conflicts with the requirement that four dimensional field theory be a good approximation to the dynamics of the inflaton. The term {\it cosmological moduli} is, within the HST formalism, an oxymoron.

\subsection{Flux compactifications}

There has been a lot of interest over the past decade in compactifications of string theory characterized by fluxes of $p$-form gauge fields through non-trivial $p$-cycles of the compactification manifold.  We would like to conjecture that the corresponding HST compactification is obtained by replacing the Dirac operator by the {\it flux Dirac operator}  $$D_F = D + \sum F_i^{(p)} \Gamma_p ,$$ where $F_i^{(p)}$ are the fluxes and $\Gamma_p$ the antisymmetrized product of Dirac matrices, contracted into the vielbein.  Spinors that are covariantly constant with respect to a generalized connection, depending on the fluxes, will give zero modes of $D_F$, which can be used to construct SUSY generators as above.  The qualitative features of the above discussion of spinor fuzzification, are unchanged by the addition of fluxes.  More quantitative details of this conjecture will be addressed in future work.

\section{Fuzzy spheres in any dimension}

The eigenvalues and eigensections of the Dirac operator on the $n$-sphere have been worked out, for example, in \cite{sphere}.  For $n$ even the eigenvalues are\footnote{We have multiplied the Dirac operator of \cite{sphere} by $i$, to make it Hermitian.}
$$ \pm (M + \frac{n}{2}),$$ where $M$ is a non-negative integer.  The degeneracy of this eigenspace is
$$D_n (M) = \frac{2^{\frac{n}{2}} (n + M -1)!}{M! (n-1)!} .$$ The eigensections are given in terms of Jacobi polynomials.  For $n$ odd we have eigenvalues
$$\pm (M + \frac{n}{2}),$$ with degeneracy
$$\frac{2^{\frac{n-1}{2}} (n + M - 1)!}{M! (n - 1)!} .$$  In both cases, the large $M$ behavior of $\Sigma_M \equiv \sum_{m \leq M} D_n (m)$ scales like $M^n$, so an eigenvalues cutoff on $M$ combined with a finite dimensional representation of the quantum algebra of variables in the spinor bundle, will have an entropy with this scaling.   This suggests that $M$ be interpreted as proportional to the radius of the sphere in Planck units.

The maximal entropy of massless particles in a region of size $R$ in $d-1$ dimensional space, subject to the constraint that they do not collapse to form a black hole with radius $\sim R$, scales like $R^{\frac{(d-1)(d-2)}{d}}$.
Now imagine that our spinor bundle variables are arranged in a $K\times L$ matrix, with $K\sim L \sim M^{\frac{d-2}{2}} $. We again try to associate particles with blocks that are roughly $P\times P$ in size.   The entropy of the factor Hilbert space generated by just those block variables is of order $P M^{\frac{d-2}{2}}.$  Thus if $P \sim M^{\frac{(d-2)^2}{2d}}$ and $M \sim R$ in Planck units, we reproduce the particle entropy formula coming from black hole physics.  The formula for $P$ shows that $P^2$ can be interpreted as the dimension of the fuzzy spinor bundle on the $d-2$ sphere, with eigenvalue cutoff $M^* \sim M^{\frac{d-2}{d}}$.  This generalizes the $M^{\frac{1}{2}}$ cutoff found in \cite{bfm}.  Following that reference we interpret this as the cutoff on the size of the longitudinal momentum $p (1, \Omega)$ in units of the inverse radius $\frac {1}{ M M_P}$ of the causal diamond.

In the four dimensions the individual $K$, $L$, or $P$ dimensional factor spaces carry irreducible representations of the rotation group.  We have not found an analog of that factorization for general $d$.   However, the formalism is completely rotation invariant, because the spaces of $K \times L$ and (roughly) $P \times P$ matrices are all spinor bundles with an eigenvalue cutoff for the Dirac operator.  Thus, the variables of our quantum theory, both the full causal diamond algebra, and the sub-algebra that describes particle-like excitations, transform as representations of the Rotation group ${\rm Spin} (d-1)$.   

For dimension greater than two, our construction is completely different from alternative approaches to the construction of fuzzy spheres (e.g.~\cite{fuzzynspheres1, fuzzynspheres2, fuzzynspheres3}).  In particular, it is rotation invariant in any dimension. We think that the origin of this discrepancy has to do with the philosophy, which has hitherto guided studies of fuzzy geometry.  This philosophy is driven by the functorial equivalence between ordinary spaces and their commutative algebra of functions.  The idea is then to take a sequence of finite dimensional non-commutative or perhaps non-associative, algebras, which, as the dimension grows large, approaches the commutative algebra of functions on some space.  One is then left with a Poisson structure, or some other tensor related to non-associativity, which encodes the leading deviation from commutative associative algebra in the large dimension limit.  This tensor is not rotation invariant on a general $n$ sphere.   In our approach, the relevant algebra is the algebra of matrices in the finite dimensional spinor bundle, with Dirac eigenvalue cutoff.  This algebra approaches (with appropriate asymptotic conditions on the matrices), the algebra of bounded operators on the Hilbert space of square integrable sections of the spinor bundle, as the eigenvalue cutoff goes to infinity.  The algebra of functions is a proper, commutative sub-algebra of this.   It is clear that there is more than enough information in this algebra to completely determine the geometry of the manifold, but our picture does not fit into the general framework of deformation quantization.   It is clear that for the purposes of HST our definition of fuzzy geometry is more suitable than others.  It remains to be seen whether it will have more general applicability.

\section{Applications to Matrix Theory}

Matrix theory is an approach to a non-perturbative construction of certain super-Poincare invariant models of string/M theory.  It should be thought of as a discrete light-cone quantization (DLCQ) of the underlying theory, in which only particle states with discrete, positive longitudinal momenta are kept and the total longitudinal momentum is restricted to be a positive integer $N$.  An elegant derivation of the Matrix Theory prescription from perturbative string theory has been given in \cite{seiberg}, following work of \cite{sen, susskind}.  One realizes the compact null direction as an infinite boost of a small space-like circle and uses the duality between M-theory and IIA string theory to claim that the positive momentum states are all D0-branes.  The light front theory needs the \textit{non}-relativistic D0 brane action, and with enough SUSY, this is completely determined.  For four or fewer compact dimensions, preserving at least 16 supercharges, the resulting theory is a well-defined quantum field theory.  From the string theory point of view, the dimensions of the compact space are small in string units, if they are $O(1)$ in $11$ dimensional Planck units, so we must do T-duality transformations (Fourier-Mukhai transformations in the case of $K3$ manifolds), to get to a frame where the physics is well understood.

For five dimensions one has to deal with the poorly understood Little String Theories\cite{LST} and for six or more compact dimensions that dual theory appears to require quantum gravity and does not achieve the objective of reducing quantum gravity to a non-gravitational problem.  One of the present authors (TB) has emphasized before\cite{matrev} that, although the Seiberg prescription is elegant and allows us to use results of perturbative string theory, there is no such thing as a {\it unique} DLCQ of M-theory.   DLCQ is an approximation method, and {\it any} approximation that gets the right results in the $N \rightarrow\infty$ limit is acceptable.
Fuzzy geometry \cite{steinacker} will enable us to define M-theory for all supersymmetric compactifications in terms of the large $N$ limit of a finite matrix quantum mechanics.

The matrix Lagrangian for Matrix Theory in 11 non-compact dimensions is

$$L = {\rm Tr}\  \bigl[\frac{1}{2R} \dot{X}^2  - \theta^T \dot{\theta} - \frac{R}{4} [X^i , X^j ]^2 - R\theta^T \gamma_i [\theta , X^i ] \bigr] .$$
$X^i$ is a $9 = 11-2$ dimensional real transverse vector of $N \times N$ matrices and $\theta$ is an $N\times N$ matrix of $16$ component real Spin$(9)$ spinors, on which the Dirac matrices $\gamma_i$ act in the usual fashion.  The $U(N)$ symmetry of the Lagrangian is a gauge symmetry, and the Super-Galilean group of the light front frame acts on the gauge invariant subspace of the the Hilbert space of this theory. The Lagrangian is written in 11 dimensional Planck units and the dimensionless parameter $R$ is the radius in Planck units of the null longitudinal circle, which determines the quantization of longitudinal momentum.  The Hamiltonian is simply proportional to $R$. In these units, the total momentum is $N$.  The claim is that as $N$ and $R$ go to infinity at fixed ratio, the states which remain at finite energy are simply supergravitons in flat 11 dimensional space-time, and the scattering matrix of those excitations along the flat directions of the quantum potential approaches the S-matrix of 11 dimensional quantum supergravity, for all momenta.

When we compactify Matrix Theory on a torus, following Seiberg's prescription, the $X^I$ for the compact directions become covariant derivatives in a $U(N)$ gauge theory on the T-dual torus. $\theta$, for each value of the non-compact spinor index, becomes a section of the tensor product of the spinor bundle over the T-dual torus, with the principal $U(N)$ bundle. Similarly, when we compactify Matrix Theory on a $K3$ manifold, four of the $X^I$ are replaced by covariant derivatives on the Fourier-Mukhai dual $\widetilde{K3}$.  The result is the $U(N) (2,0)$ six dimensional CFT, which is the unique maximally supersymmetric UV completion of $5$ dimensional SYM, compactified on $S^1$ times $\widetilde{K3}$\cite{rozali}.

Our proposal for Matrix Theory compactification is to take the original  Seiberg proposal, which naively takes the theory to a SYM theory on the dual of the compactification manifold, and replace that manifold by its spinor fuzzification.  Thus, the $X^I$ become covariant derivative operators in a bundle of $N^2$ spinor fields over the manifold\footnote{In the Matrix Theory Lagrangian, we recognize that the compact $X^I$ are the {\it tangent space components} of covariant derivatives, $e^{\mu}_a D_{\mu}$, so that the flat scalar product is all that is necessary.}, with a cutoff on the Dirac eigenvalue that is related to the size of the dual compactification manifold in Planck units.  Each non-compact component of $\theta_a$ is a section of this bundle.  Each non-compact $X^I$ is a function on this manifold.  That is to say, it is in the tensor product of the algebra of $N\times N$ matrices, with the 0-form subalgebra of the Clifford algebra of forms on the fuzzy manifold.

According to this proposal, Matrix Theory compactification in {\it any dimension} is a supersymmetric quantum mechanics of finite dimensional matrices.   The only issue one has to deal with is whether the large $N$ limit (with the size of the compact spinor bundle fixed) defines a finite, super-Poincare invariant S-matrix.   This prescription is even applicable to $G2$ compactification, or compactification on a $7$-torus.   Indeed, it even allows us to define a finite $N$ version of compactification on $8$ or $9$ dimensional manifolds.  Presumably, in those cases, the large $N$ limit of the scattering matrix fails to exist.   We hope to come back to some examples of finite Matrix Theory compactifications in a future publication.

One question left open by this proposal is what we mean by ``dual" in the general case.  For tori and K3 manifolds this is clear.  The authors of \cite{kln} suggested that for $CY3$-folds the relevant duality is mirror symmetry.  Indeed, the problem Seiberg solved with T-duality was that the description in terms of $D0$-branes on the original manifold had a manifold whose size shrank to zero in string units.   The string perturbation expansion breaks down, and sometimes the duality tells us how to describe the resulting limit in an exact way.   The mirror dual of a zero volume $CY3$-fold is a $CY3$-fold at its conifold singularity.
In fact, our discussion of fuzzy compactification and the holographic principle suggests that when the size of the manifold is of order 1 in Planck units, the approximation of continuous moduli breaks down.  The manifold and it's mirror dual are just two different $O(1)$ values of the discrete moduli.   

Greg Moore has suggested to us a strategy which would obviate the need for formulating a precise notion of dual to every compactification manifold.  In all known examples, the Dirac-Ramond operator, with supersymmetric boundary conditions, has a spectrum invariant under dualities of string theory that preserve $g_S = 0$.  An eigenvalue cutoff on the Dirac-Ramond operator, again leads to a finite dimensional spinor bundle, so perhaps this could be used as a definition of fuzzy compactifications of Matrix Theory.

\section{Conclusions}

The Strong Holographic Principle implies that a finite area holographic screen corresponds to a finite dimensional approximation to the spinor bundle over the screen.  Defining this approximation by a sharp cutoff on the spectrum of the Dirac operator, preserves all isometries of the manifold, as well as SUSY.  It gives a rather precise definition of compactifications of the holographic space-time formalism, as well as compactifications of Matrix Theory.  The latter always correspond to a quantum system with a finite number of variables.  The only question that arises is whether the large $N$ limit of the Matrix Theory scattering matrix converges to a Super-Poincare invariant answer.
\vspace{2pc}

\noindent
{\bf  Acknowledgements:}
The authors would like to acknowledge useful conversations with Greg
Moore.  This work was supported in part by the U.S.~Department of
Energy.



\providecommand{\href}[2]{#2}\begingroup\raggedright\endgroup


\end{document}